\documentclass[review]{elsarticle}

\usepackage{lineno,hyperref}
\usepackage{amsmath,amsthm}
\usepackage{amssymb}
\usepackage{graphicx}
\usepackage{tikz}
\usepackage{geometry}
\modulolinenumbers[5]

\begin{document}

\begin{frontmatter}

\title{Modelling the behavior of human crowds as coupled \\  active-passive dynamics of interacting particle systems}
\tnotetext[mytitlenote]{}

\author[mymainaddress,myaddress]{Thoa Thieu \corref{mycorrespondingauthor}}
\cortext[mycorrespondingauthor]{Corresponding author}

\author[mymainaddress,mysecondaryaddress]{Roderick Melnik}


\address[mymainaddress]{M3AI Laboratory, MS2Discovery Interdisciplinary Research Institute, Wilfrid Laurier University, 75 University Ave W, Waterloo, Ontario, Canada N2L 3C5}
\address[myaddress]{Department of Mathematics, Indiana University, 831 East 3rd St., Bloomington, IN 47405}
\address[mysecondaryaddress]{BCAM - Basque Center for Applied Mathematics, Bilbao, Spain \\Emails: tthieu@iu.edu, rmelnik@wlu.ca}

\begin{abstract}
The modelling of human crowd behaviors offers many challenging questions to science in general. Specifically, the social human behavior consists of many physiological and psychological processes which are still largely unknown. 
To model reliably such human crowd systems with complex social interactions, stochastic tools play an important role for the setting of mathematical formulations of the problems. In this work, using the description based on an exclusion principle, we study a statistical-mechanics-based lattice gas model for active-passive population dynamics with an application to human crowd behaviors. We provide representative numerical examples for the evacuation dynamics of human crowds, where the main focus in our considerations is given to an interacting particle system of active and passive human groups. Furthermore, our numerical results show that the communication between active and passive humans strongly influences the evacuation time of the whole population even when the \textquotedblleft faster-is-slower\textquotedblright \ phenomenon is taken into account. To provide an additional inside into the problem, a stationary state of our model is analyzed via current representations and heat map techniques. Finally, future extensions of the proposed models are discussed in the context of coupled data-driven modelling of human crowds and traffic flows, vital for the design strategies in developing intelligent transportation systems.
\end{abstract}

\begin{keyword}
Crowd dynamics\sep Collective behavior \sep Social group interactions \sep Faster-is-slower \sep Particles current \sep Kinetic Monte Carlo and Markov chains \sep Occupation numbers \sep Evacuation dynamics \sep  Simple exclusion dynamics \sep Lattice Boltzmann and lattice gas cellular automata methods \sep Social networks
\end{keyword}

\end{frontmatter}


\section{Introduction}

Social dynamics and human behavior are closely connected. Moreover, the detailed
behavior of human crowds is already complicated, being caused by many physiological, sociological
and psychological processes, and it is still largely unknown. The topic of human crowds offers many challenging questions to the scientific community. In crowd dynamics studies, the movements of human groups must directly be related
to their decision making processes. Hence, the characteristics of human flows
are apparently affected by decisions of each individual in the group. This influence needs to be accounted for
in order to model the processes properly. In an emergency situation, in particular in urban areas \cite{Bretschneider2013,Shin2019,Kubera2021,Ronchi2021}, individuals require information about surrounding environment and social interactions in order to evacuate successfully. Therefore, stochastic models and tools become important to
capture the essence of the interactions among such individuals (e.g. \cite{Cirillo2019,Cirillo2020,Thieu2021}). One of the most promising routes in this direction is the development of statistical-mechanics-based lattice gas system models for the dynamics of active-passive populations. In general, contributions from psychology are also needed to model the dynamics of
emotional behaviors which can affect the human strategy. Likewise, contributions from
neuroscience may also be vital to better understand the collective learning dynamics of people that face
emergency situations \cite{Pelechano2005,Templeton2018}. By now we know that the emergence of collective opinions in a multi-agent system brings about a challenging situation when it comes to modelling complex biosocial systems \cite{Tadic2020, Tadic2021}. The formation
of (partial) consensus among agents can lead to such collective opinions in many
cases \cite{Spreen2020}. There are a number of relevant results available on the topics of the behavior of human crowds. In particular, a
relevant result on dissonance minimization as a micro foundation of social influence
in models of opinion formation has been reported in \cite{Groeber2014}, where the authors show
that different models of opinion formation can be represented as best response
dynamics within a general framework of social influence. A model for dynamics of conflicting opinions considering rationality has been studied in \cite{Wang2020}. In \cite{Lee2020}, the authors experimentally investigated the effect of different proportions of patient (no-rush) versus impatient (rush) individuals in an evacuating crowd. A multiscale vision of human crowd models, which provides a consistent description at the three possible modelling scales, namely, microscopic, mesoscopic, and
macroscopic, has been reported in \cite{Aylaj2020}. At the application level, the authors of \cite{Bellomo2020} developed a model within a multiscale framework accounting for the interaction of different spatial scales, from the small scale of the SARS-CoV-2 virus itself and cells, to the large scale of individuals and further up to the collective behaviour of populations. One of the most common phenomena in crowd dynamics is the phenomenon of self-organization. It is worth pointing out that human crowds are able to produce coherent flow patterns, and demonstrate manifestations of certain built-in self-organization mechanisms.  For example, as shown in \cite{Helbing2005, Cristiani2015}, pedestrian flows, either uniform or not, can form collective patterns of motion. For instance, one notices circulating flows at intersections, crowd flows at bottlenecks, lane formations, local cloggings due to a complex geometry \cite{Schadschneider2011, Tordeux2020}. These behaviors are typically caused by walls and obstacles under normal walking conditions, including the situations when the evacuation of pedestrians takes place during an emergency situation. Furthermore, the well-known \textquotedblleft faster-is-slower\textquotedblright \ effect is an important instance of self-organized phenomenon in crowd dynamics \cite{Oh2017, Kruchten2017, Garcimartin2014}. In general, the \textquotedblleft faster-is-slower\textquotedblright \ effect is caused by the impatience of human groups with panic moods in an emergency situation. Another example is provided by the clogging of humans at an exit door or a bottleneck that could lead to fatal accidents. A better understanding of the \textquotedblleft faster-is-slower\textquotedblright \ phenomenon would help to avoid unexpected accidents and reduce the evacuation time of human groups in emergency scenarios.

In this paper, we study an interacting particle system, modelling the dynamics of a mixed group of active-passive particles with applications to human crowd behaviors. We provide representative numerical examples of an evacuation dynamic of active-passive human groups. In particular, we focus on the situation where an active human group is aware of
the details of the environment and move towards the exit door, while a passive population is not aware of the details of the geometry and move
randomly to explore the environment and eventually to find the exit.
In this context, humans interact by communicating between active and passive human groups. Specifically, the active human group transfers the information of surrounding environment to the passive human group. We give our special attention to representative numerical examples on the evacuation dynamics of such active-passive human groups, where we observe a classical \textquotedblleft faster-is-slower\textquotedblright \ effect. Furthermore, a stationary state of our model is also investigated, with corresponding current and heat map characteristics discussed in detail.

From a larger perspective of social group interactions, an increasing interest in the literature is
devoted to active-passive and prosocial-antisocial coping strategies \cite{Chwaszcz2022}. Such social group
interactions are fundamental to our better understanding of active-passive human groups and
to developing efficient strategies to address a growing number of social problems, such as
social exclusion. In our opinion, a cornerstone to this is advancing the models that can predict
the human crowd dynamics in critical situations and cases of emergency, such as fire
accidents, terrorist attacks, and power failures that require evacuations. That is why our main
exemplifications in this work come from that area. Further, the resulting models can provide
additional insight into and improve our understanding of human decision-making processes
under uncertainty \cite{Liu2022,Thieu2023}.

The models developed here are also useful for the analysis of social networks, including those
with computer-mediated communication, where uncertainty can be reduced based on active,
passive, or interactive strategies of responses, as well as for the study of emergent collective
knowledge in the dynamics of meaningful social interactions \cite{Antheunis2010,Dankulov2015,Andjelkovic2016,Tadic2017,Tadic2021self,Huang2022}. These areas of social
group interactions and social networks, where the active-passive strategies are essential, can
particularly benefit from the models developed here. Given their relative simplicity and
computational efficiency, they have clear advantages compared to earlier developed models
and existing methods. Indeed, the problem at hand is a multiscale many-body problem for
interacting (communicating) particles, which is notoriously difficult to solve in its generality.
Viable approximations have to be used depending on the scale of consideration and details to
be accounted for. Our focus in the forthcoming sections will be on developing novel
mesoscopic models, accounting for the effect of communication between active-passive
populations in the crowd, with a rigorous foundation rooted in lattice Boltzmann methods and
their historical precursor, gas lattice cellular automata.

\section{Developing models for communicating active-passive populations in human crowds}\label{model}
An essential feature of human crowds that is often overlooked in the traditional modelling
approaches in this field is the communication between individual agents, which is modulated
by a broad spectrum of psychological effects, especially pronounced in extreme situations. In
what follows, we make the first steps in this direction by analyzing the impact of
communication between active-passive human groups.
Our approach originates in the well-known lattice Boltzmann method (LBM). Its
mathematical foundation is linked to conservation laws for macroscopic properties (e.g., mass,
momentum, and energy for the Navier–Stokes equations or energy-balance models of
hydrodynamic type), assuming that the matter under consideration behaves according to the
Newton laws (although modifications can be made to account also for more complex
behaviours). This versatile, dynamic method, which has been applied extensively to various
fluid dynamics problems, can successfully simulate the macroscopic behaviour of fluids using
much simpler mesoscopic models. Unlike computational fluid dynamics methods (which
numerically solve the conservation equations of macroscopic properties), the matter in LBM
models is represented by particles that consecutively move along a discrete lattice (with
streaming/propagation and collision/relaxation processes). This representation leads to a
number of advantages of LBM models, including reasonably simple algorithms for their
implementation, easy programming of LBM-based parallel tasks, efficient management of
complex geometries, and possibilities of simulating various complicated nonlinear
macroscopic phenomena. In fact, from a macroscopic point of view, these models allow us to
connect macroscopic and microscopic scales of multiscale dynamics of complex matter.
Importantly, in their generality, these discrete models are based on and start from nonequilibrium
statistical mechanics. Hence, many authors begin their consideration from the
Boltzmann equation (e.g., \cite{Bellomo2022towards,Thieu2023review,Bellomo2023human,Corbetta2023} and references therein), which, in this case, can be
interpreted as a discrete-velocity Boltzmann-type model. As seen in these references, this
kinetic-theory-inspired consideration is now also widely used in modelling human crowds at
the mesoscopic scale. With the evolution of probabilities described by the Boltzmann
equation instead of describing gas molecules individually, the kinetic theory was originally
developed for gases. In this context, it should also be mentioned that the LBM model itself
originates from the lattice gas cellular automata (LGCA), and some of the standard algorithms
implementing these ideas lead to HPP (Hardy-Pomeau-Pazzis), FHP (Frisch-Hasslacher-
Pomeau models), and MZ (McNamara and Zanetti) models, among many others.
In addition to cellular automata models, there is a range of other tools, including path integral,
Monte Carlo and Markov chain methods, their combinations, multigrid, and other approaches,
that are relevant to this scale of human crowd description (e.g., \cite{Corbetta2023,Tamang2023,Xu2023}). Despite
LBM advantages, these models are very memory-intensive due to the necessity to store the
distribution functions, leading to one of the main bottlenecks of LBM. At the same time, the
macroscopic level of description of crowd dynamics based on continuum-mechanics
equations (including, but not limited to, the Navier-Stokes equations) is inadequate practically
in all cases where biosocial interactions of agents become essential (and the recent pandemic
provides an instructive example of such situations, e.g. \cite{Tadic2020,Tadic2021}). On the
other hand, microscopic scale details that usually require Molecular Dynamics simulations
based on Hamiltonian dynamics models may be prohibitively computationally expensive and,
hence, frequently unfeasible. We also note an interesting recent development in microscopic
theories based on closure approximations that can do more in describing complex dynamics
than standard mean-field models \cite{Rizkallah2022,Rizkallah2023}. This development could be helpful for some of
the problems considered here and motivate their further extensions once such theories are
fully established. Based on the above, we conclude that the LGCA-type models provide a
good compromise between efficiency and computational complexity. Therefore, these models
will be taken as a basis for further development here. The application of these models in
attempts to describe human behaviour is not new, and such attempts have included evacuation
processes for pedestrian dynamics accounting for some active and passive factors (e.g., \cite{Song2006}
and references therein). Further, cellular automata models have been extended to the
description of the behaviour of human crowds in extreme situations, such as terrorist attacks
\cite{Song2022}. However, in no way, even these most recent works included active-passive population
dynamics and, what is particularly important, the effect of communication between activepassive
human groups. It brings clear evidence of our models’ strengths and innovations
compared to the current state-of-the-art in the field. LGCA models are fundamental in many
branches of science and engineering. Yet, their applicability to human crowds involving
social and biosocial dynamics has not achieved the level of development it deserves.
\subsection{Lattice gas}
A lattice gas is a set of particles jumping at random on a lattice (see, e.g., in \cite{Thieu2020thesis,Spohn1991}). We suppose that the positions of particles are restricted to a discrete subset of $\mathbb{R}^d (d \in \mathbb{N}^{\ast})$. In general, this subset is taken to be the d-dimensional cubic lattice
\begin{align}
	\mathbb{Z}^d := \left\{i=(i_1, \ldots, i_d) \in \mathbb{R}^d: i_k \in \mathbb{Z} \text{ for each } k \in \{1, \ldots, d\} \right\}.
\end{align}

Let a lattice $\Lambda = \mathbb{Z}^d$, and a state space for the lattice gas is a configuration $\eta \in \Omega= \{0,1\}^{\Lambda}$.  Since the dynamics of the lattice gas consists of a sequence of jumps. To keep track of the number of particles, let us introduce the so-called occupation number $\eta_i$ of particles at lattice site $i$. In addition, $\eta_i$ takes values in $\{0,1\}$ such that
\begin{align}
	\eta_i = \begin{cases}
		0 &\text{ if site }i \text{ is vacant},\\
		1 &\text{ if site }i \text{ is occupied}.
	\end{cases}
\end{align}


Given the rates, we construct lattice gas dynamics in the standard fashion. We consider a finite volume $\Lambda$, with $|\Lambda| < \infty$, where $|\Lambda|$ represents the number of points in the lattice $\Lambda$. Hence, the state space consists of a finite number of configurations. Therefore, the generator of the Markov jump process acts on a function $f: \Omega \longrightarrow \mathbb{R}$ is defined by 
\begin{align*}
	\mathcal{L}(\eta) = \sum_{\eta' \in \Omega}c(\eta, \eta')[f(\eta') - f(\eta)],
\end{align*}
where $c(\eta, \eta')$ represents the rate of exchange of the
occupancies from the configuration $\eta$ to $\eta'$.
\subsection{Model description}
Our starting point is a crowd dynamics model for active-passive particles. This particle system is considered within the geometry confined to a square lattice $\Lambda:=\{1,\dots,L\}\times\{1,\dots,L\}\subset \mathbb{Z}^2$ of side 
length $L$, where $L$ is an odd positive integer number. In the examples that follow the lattice $\Lambda$ can be interpreted as a {\em room}, where  
a point $x=(x_1,x_2)$ of the room $\Lambda$ is called a \emph{site}. When
two sites $x,y$ are at the Euclidean distance one, i.e. $|x-y|=1$, we call them \emph{nearest neighbors}. The \textquotedblleft door\textquotedblright  \ is represented as a set made of $w_{\mathrm{ex}} = \omega$ pairwise adjacent sites, with $\omega < L$ being an odd positive integer, located on the top row of the room $\Lambda$ and symmetric with respect to its median column. This representation of the door can be considered as an exit door with its \textit{width} $\omega$ of the exit. We denote by $N_A$ the total number of active particles, and by $N_P$ the total number of passive particles with $N:=N_A+N_P$ and $N_A, N_P, N \in \mathbb{N}$.

\begin{figure}[h!]
	\centering
	\vspace*{-10mm}
	\includegraphics[width=0.45\textwidth]{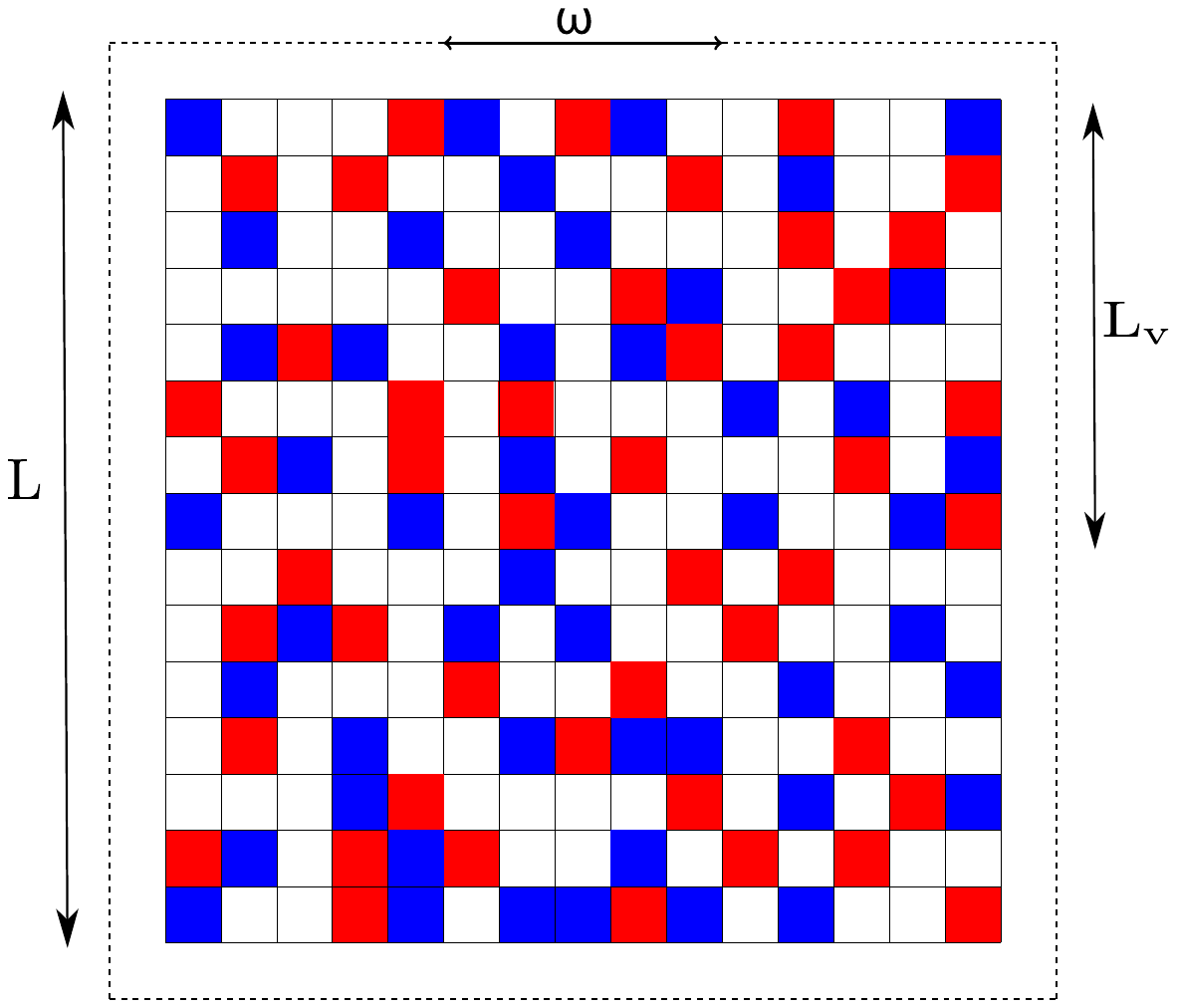}
	\vspace*{-18mm}
	\caption{(Color online) Schematic representation of our lattice model with size $L$. Blue and red 
		squares represent passive and active particles, respectively, while the white squares within the geometry denote the empty spots. The reflecting boundary conditions is represented by the thick dashed line surrounding a large
		fraction of the grid. An exit door is located in presence of the arrow, with its width equal to $\omega$. The visibility region with the length $L_{\text{v}} \leq L$ is represented in the geometry.}
	\label{fig:fig0}
\end{figure}

Furthermore, we define a rectangular interaction zone $V$ (\emph{visibility region}) inside the top part of the 
room $\Lambda$. This visibility zone is made of the 
first $L_{\text{v}}$ top rows of $\Lambda$, where the positive integer 
$L_{\text{v}}\le L$ called \emph{depth} of the visibility region, see in Fig. \ref{fig:fig0}. We refer to the case in which no 
visibility region is considered by writing $L_{\text{v}}=0$. 

In the remainder of this section, we highlight several key points important in further analysis. We focus on two different species of
\emph{active} and \emph{passive} particles, moving inside $\Lambda$ and we use two symbols A and P to 
respectively refer to them.
Note that particles cannot access the sites 
of the external boundary of the room, i.e. there are sites  
$x\in\mathbb{Z}^2\setminus\Lambda$ such that there exists 
$y\in\Lambda$ nearest neighbor of $x$. 
We call the occupation number $\eta_x$ and the state of the system
\emph{configuration} $\eta\in\Omega=\{-1,0,1\}^\Lambda$, where we have:
\begin{align*}
\begin{cases}
\eta_x=0, &\text{ if the site $x$ is 
	\emph{empty}},\\
\eta_x=1, &\text{ if the site $x$ is  \emph{occupied by an active particle}},\\
\eta_x=-1, &\text{ if the site $x$ is \emph{occupied by a passive particle}}.
\end{cases}
\end{align*}
Furthemore, the quantities $n_{\text{A}}(\eta)=\sum_{x\in\Lambda}\delta_{1,\eta_x}$
and $n_{\text{P}}(\eta)=\sum_{x\in\Lambda}\delta_{-1,\eta_x}$ represent the numbers of active and passive
particles in the configuration $\eta$, respectively, with $\delta_{\cdot,\cdot}$ being the Kronecker symbol.
The total number of particles in the configuration $\eta$ is the sum of $n_{\text{A}}(\eta)$ and $n_{\text{P}}(\eta)$.

The overall dynamics of our system can be represented by the continuous time Markov chain $\eta(t)$ on 
$\Omega$ with rates $c(\eta,\eta')$ defined as in \cite{Cirillo2019}. 
Markov chains provide a well-established tool for modelling complex dynamic evolution, including the situations where multiscale features of the dynamics have to be accounted for \cite{Melnik1998,Melnik2008, Melnik2009,Besenczi2021}.
Based on our Markov chain representation, we are now in a position to analyze the dynamics of our active and passive human groups. In this case, using a simple exclusion process, the underlying dynamics can be modelled as follows: the passive population performs a 
symmetric simple exclusion dynamics on the whole lattice, whereas the active population is subject to a drift, guiding particles towards the exit door.

In our model, let $\varepsilon\ge0$ be the \emph{drift}. We assume that only active particles experience the drift, 
changes of the parameters $L_{\textrm{v}}$ and $\varepsilon$
act directly only on the active species, while passive particles do not depend on the drift value. While in the general case, the drift can depend on $x,y$ and $t$, in the examples considered in Section 3, we assume it to be a constant.


The above construction follows the ideas originally discussed in \cite{Cirillo2019, Cirillo2020}. In these earlier works, the interaction of particles inside the room was modelled via a 
simple exclusion random walk for two particle species  undergoing 
two different microscopic dynamics. The passive particles performed a 
symmetric simple exclusion dynamics on the whole lattice, while the active particles performed a symmetric simple exclusion walk 
outside the visibility region, whereas inside such a region they 
experienced a drift pushing them towards the exit. The active particles were obscure only in the region outside the visibility region, while the whole lattice was obscure for the passive particles.
 In our current consideration, motivated by the recent results of \cite{Thieu2021}, we have modified this earlier model by incorporating the interaction between active and passive particles and analyzing it on an example of modelling the dynamics of a mixed group of active-passive humans. In particular, we assume that a passive human can communicate with active humans at his nearest neighbor sites. After receiving enough information from the active humans, the passive human becomes active and has the same behavior as the active one. Further details of the underlying scheme and its numerical implementation are provided in Section \ref{num}.



\section{Numerical results}\label{num}

In this section, we consider an application of our interacting particle system, modelling the dynamics of human groups. In particular, we investigate an interacting particle system of active and passive populations on the example of interacting active and passive human groups. We assume that active particles can be seen as active humans while passive particles can be considered as passive humans and both human groups wish to escape the room due to an emergency situation. The underlying dynamic process can be described as follows: the active humans have the information about the geometry (knowing the location of the exit door) and automatically transfer the information to the passive human group. After receiving the information from the active human group, the passive humans  follow the active human group. Hence, the passive humans are \textquotedblleft switched\textquotedblright \ to active humans. At this moment, we assume that the passive human group trusts the active human group by communicating with active humans at its nearest neighbor sites. Moreover, we assume that
the drift quantity $\varepsilon$ represents the level of panic mood measuring the need of escaping the room in an emergency situation.
In particular, the larger $\varepsilon$ of active humans, the more active humans wish to get out of the room.

\begin{figure}[h!]
	\begin{center}
		\resizebox{12cm}{5cm}{
			\begin{tikzpicture}[scale=1.5,every node/.style={transform shape}]
			\filldraw[fill=green!20,xshift=-5.5cm,xscale=5.5] (-1,1)--(1,1)--(1,-1)--(-1,-1)--(-1,1);
			\draw[xshift=-5.5cm] (0,0) node[align=center] {\bf Extract at time $t$ an exponential random time $\tau$  as a \\ \bf function with parameter which is 
				the total rate 
				\\\bf	$\sum_{\zeta\in\Omega}c(\eta(t),\zeta)$ of exchanging between configurations}; 
			\filldraw[fill=yellow!20!,xshift =5.5cm,xscale = 3.5] (-1,1)--(1,1)--(1,-1)--(-1,-1)--(-1,1);
			\draw[fill = green,xshift=5.5cm] (0,0) node[align=center] {\bf Set $t= t+\tau$};
			\filldraw[fill=red!10!,yshift = -5cm,xscale = 4.5] (-1,1)--(1,1)--(1,-1)--(-1,-1)--(-1,1);
			\draw[fill = green,yshift=-5cm] (0,0) node[align=center] {\bf Select a configuration using the \\ \bf probability 
				distribution 
				$c(\eta(t),\eta)/\sum_{\zeta\in\Omega}c(\eta(t),\zeta)$
				\\ \bf and set $\eta(t+\tau)=\eta$};
			
			\draw[xshift = -4.5cm,->,thick] (4.5,0)--(6.5,0) node[midway,sloped,above]{\color[HTML]{008080} };
			\draw[xshift = -4.5cm,->,thick] (4.5,0)--(6.5,0) node[midway,sloped,below]{\color[HTML]{008080} };
			\draw[xshift = 5.5cm,thick] (0,-1)--(0,-5) node[midway,sloped,above]{\color[HTML]{008080} };
			\draw[xshift = 5.5cm,thick] (0,-1)--(0,-5) node[midway,sloped,below]{\color[HTML]{008080} };
			\draw[xshift = 5.5cm,->,thick] (0,-5)--(-1,-5) node[midway,sloped,above]{};
			\end{tikzpicture}	}
	\end{center}
\caption{Key dynamic steps in the KMC numerical scheme implementation based on the continuous Markov chain $\eta(t)$.}\label{fig:diagram}
\end{figure}
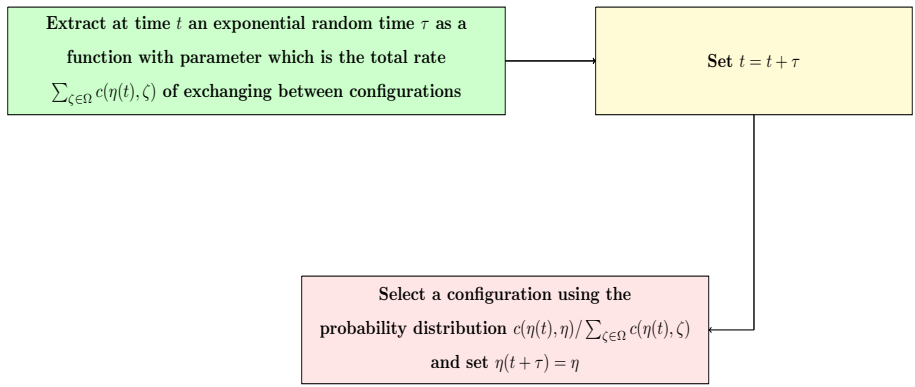

The numerical results reported in this section are obtained by using the kinetic Monte Carlo (KMC) method. The methodology is effectively based on a dynamic Monte Carlo technique and follows in its essence \cite{Cirillo2019}, where a lattice-gas-type version of KMC was applied to the study of pedestrian flows. In particular,
we use the scheme in Fig. \ref{fig:diagram} to
simulate the presented model.
Note that as in all KMC versions, we need to define the rate $c(\eta(t),\eta)$, which is done here in the same manner as in \cite{Cirillo2019}).

\begin{figure}[h!]
	\centering
	\vspace*{-10mm}
	\includegraphics[width=.3\linewidth]{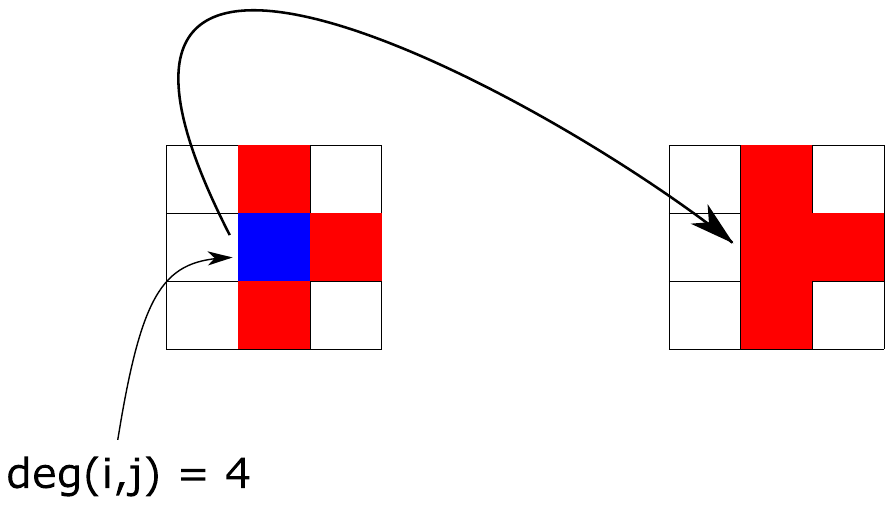}\hspace{0.01cm}
	\includegraphics[width=.3\linewidth]{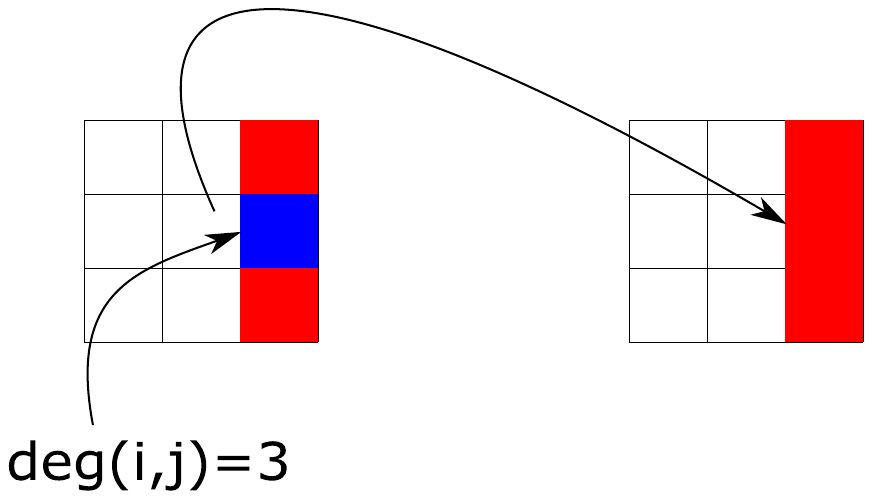}\hspace{0.01cm}
	\includegraphics[width=.3\linewidth]{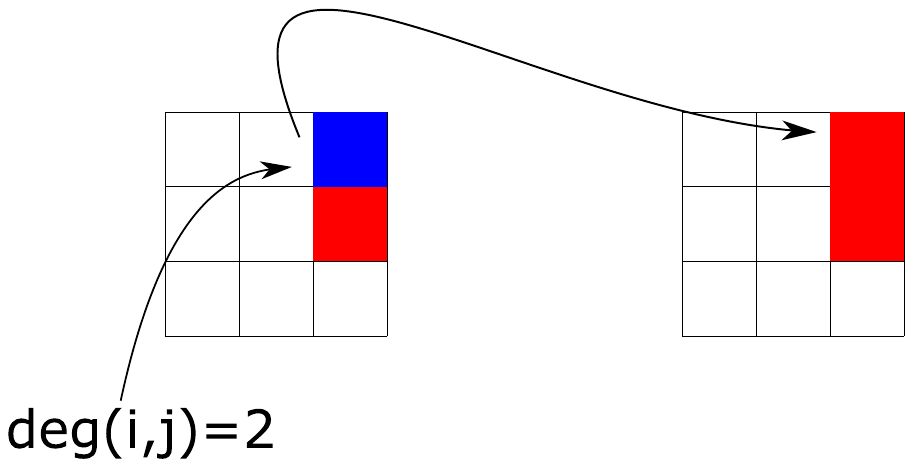}
	\vspace*{-25mm}
	\caption{\small (Color online) Schematic representation of the interaction between active and passive particles.}
	\label{fig:fig0-1}
\end{figure}
\begin{figure}[h!]
	\centering
	\begin{tabular}{lll}
		\includegraphics[width=0.3\textwidth]{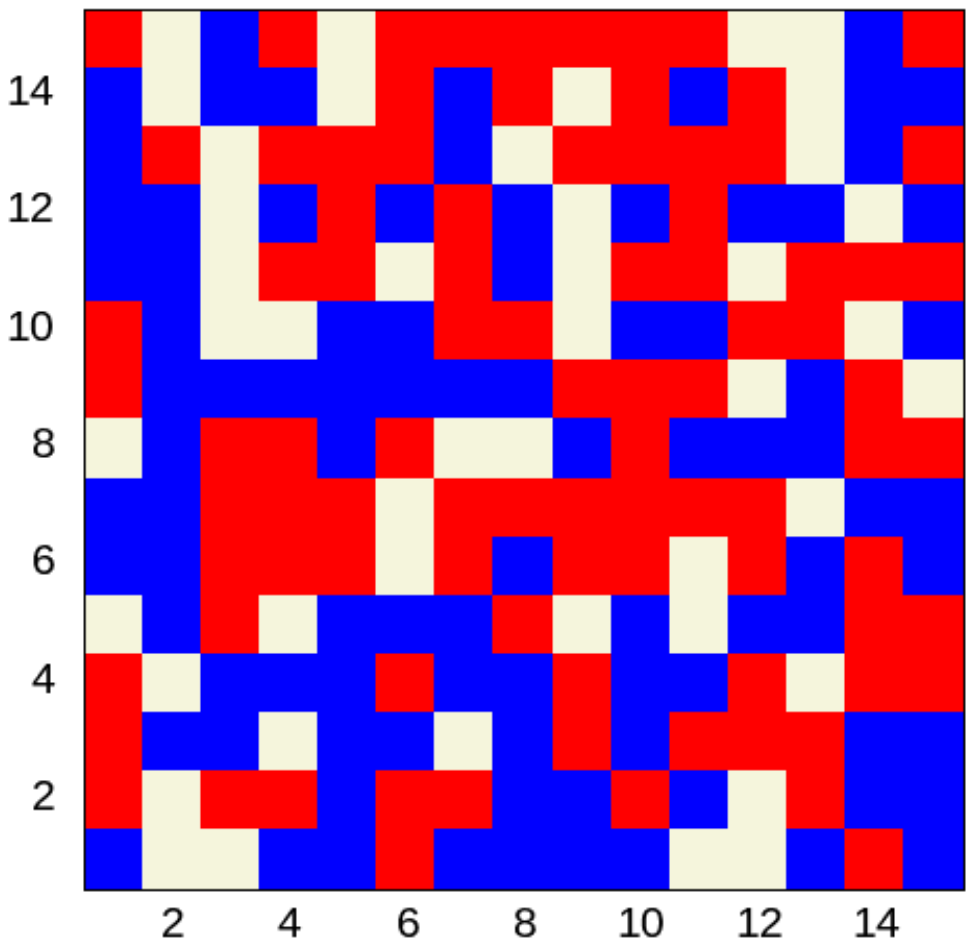}&
		\includegraphics[width=0.3\textwidth]{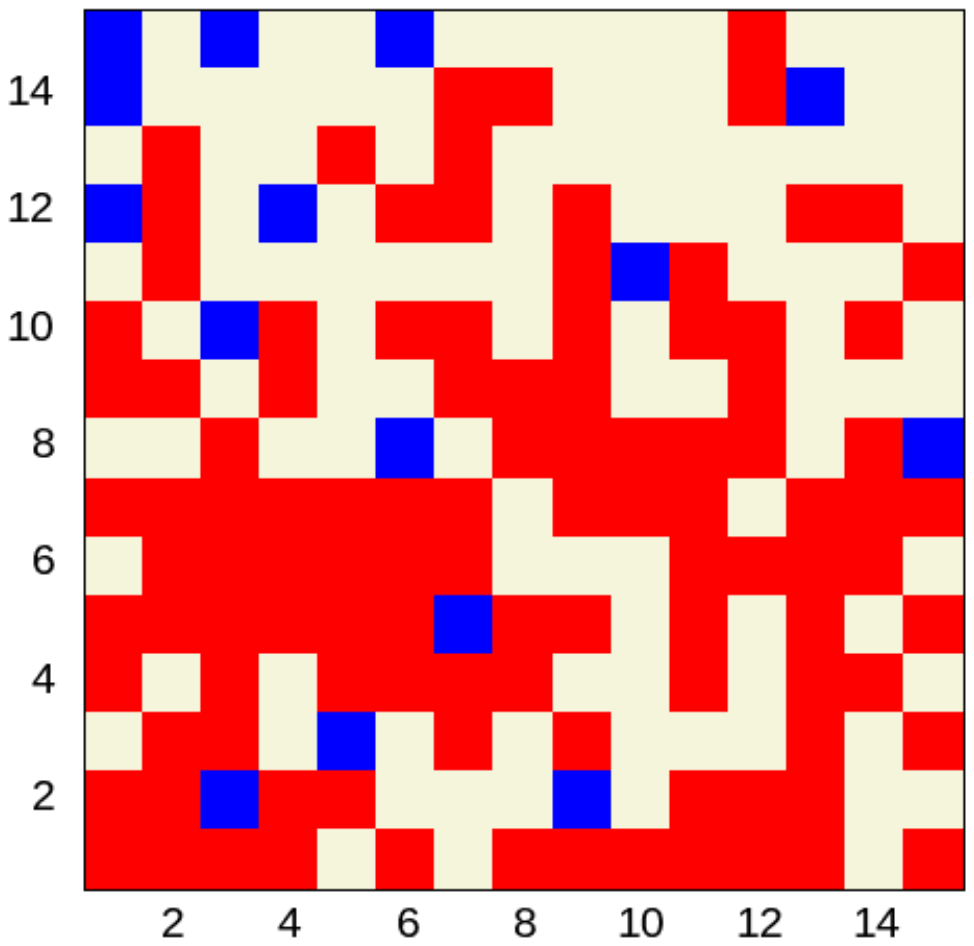}&
		\includegraphics[width=0.3\textwidth]{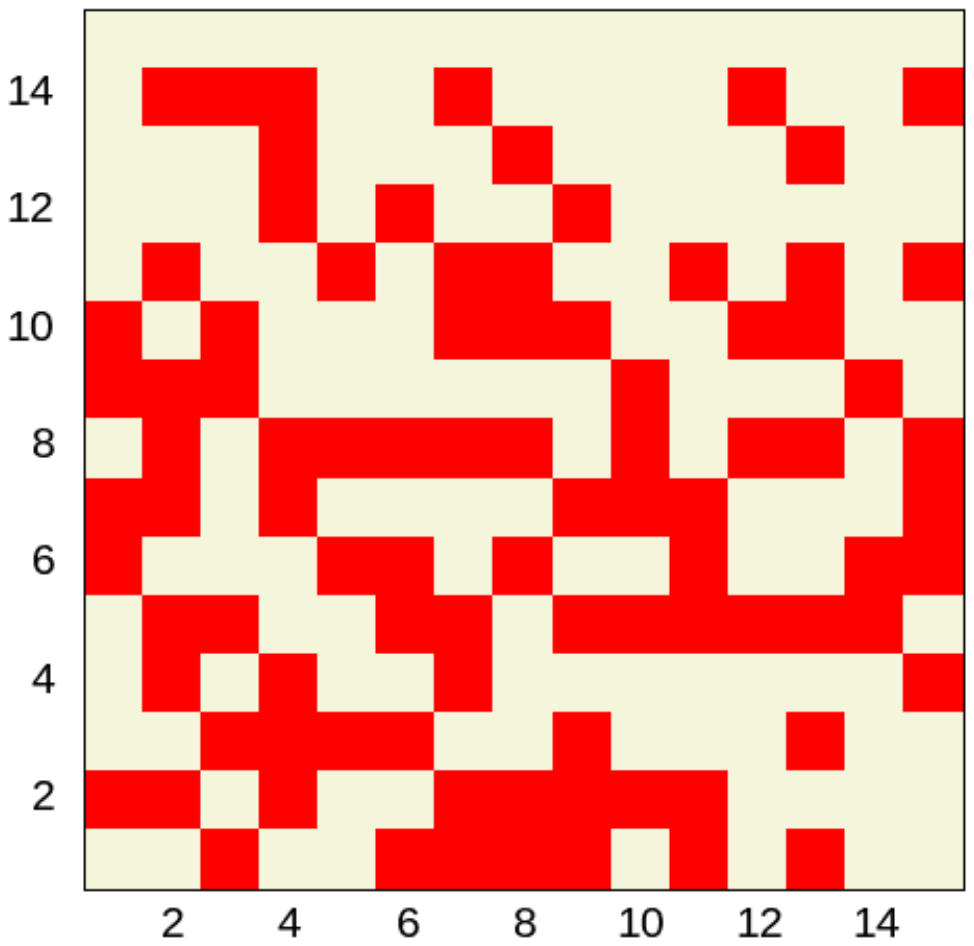}\\[0.1cm]
		\includegraphics[width=0.3\textwidth]{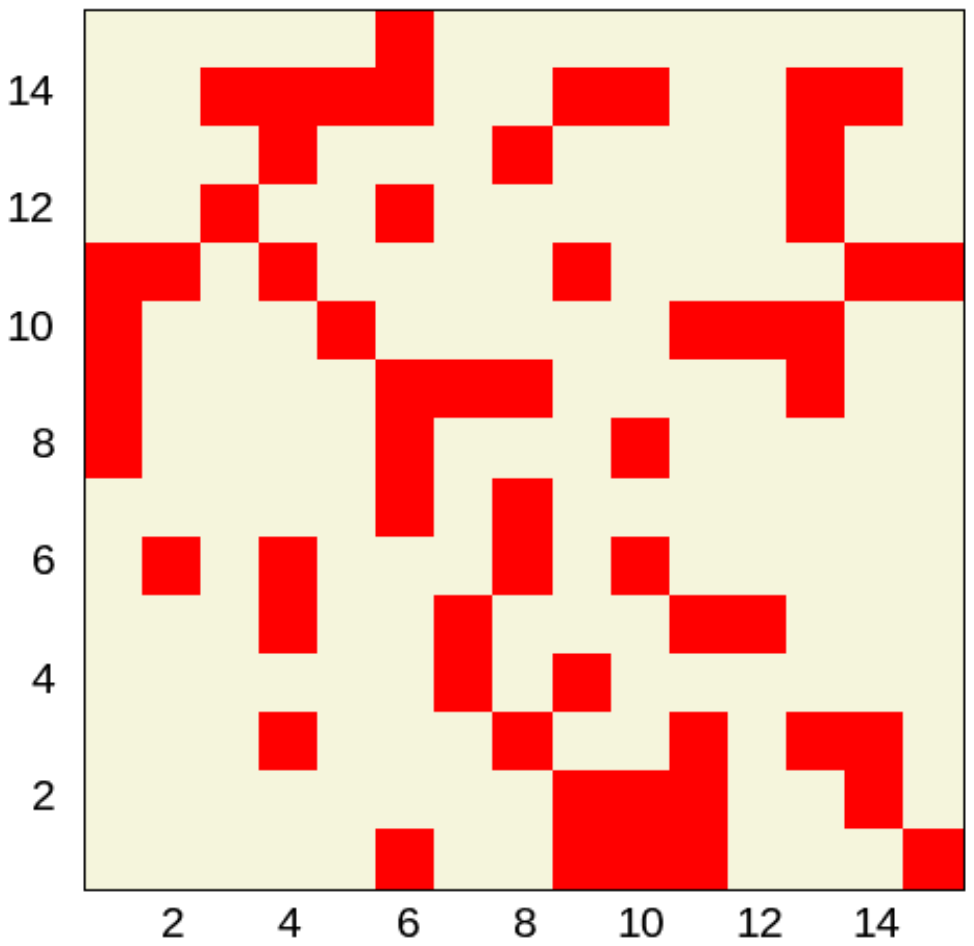}&
		\includegraphics[width=0.3\textwidth]{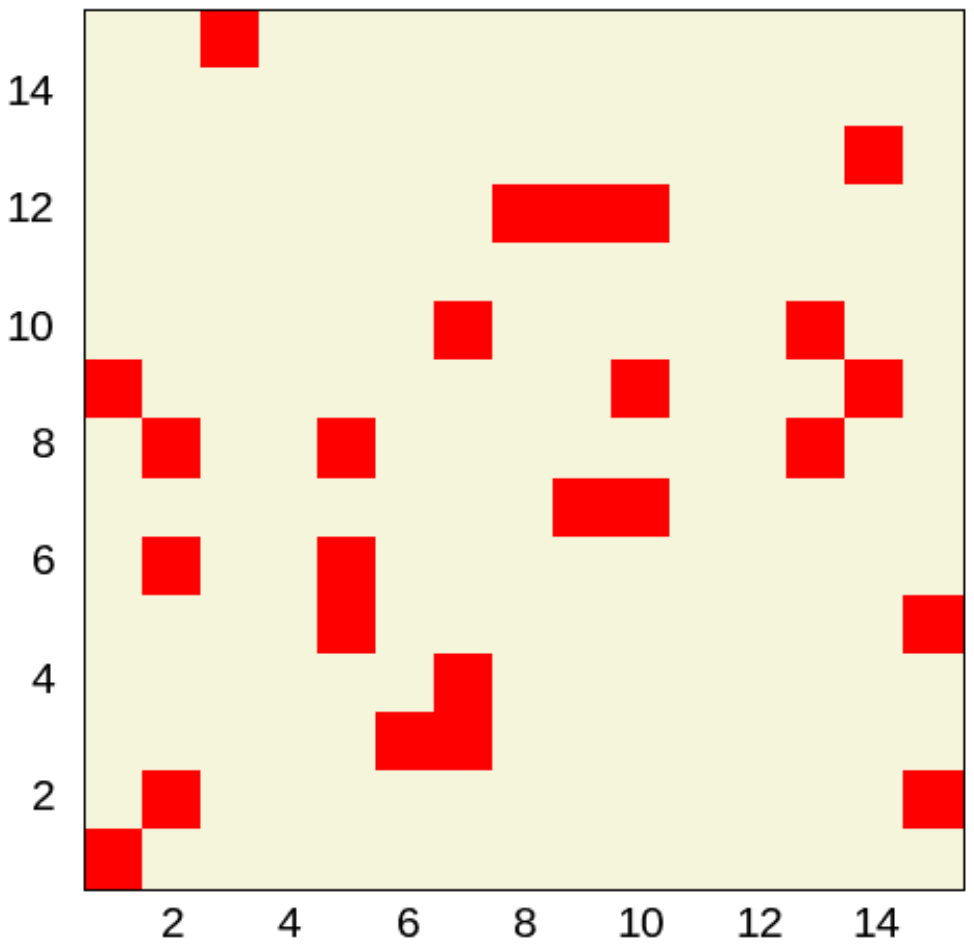}&
		\includegraphics[width=0.3\textwidth]{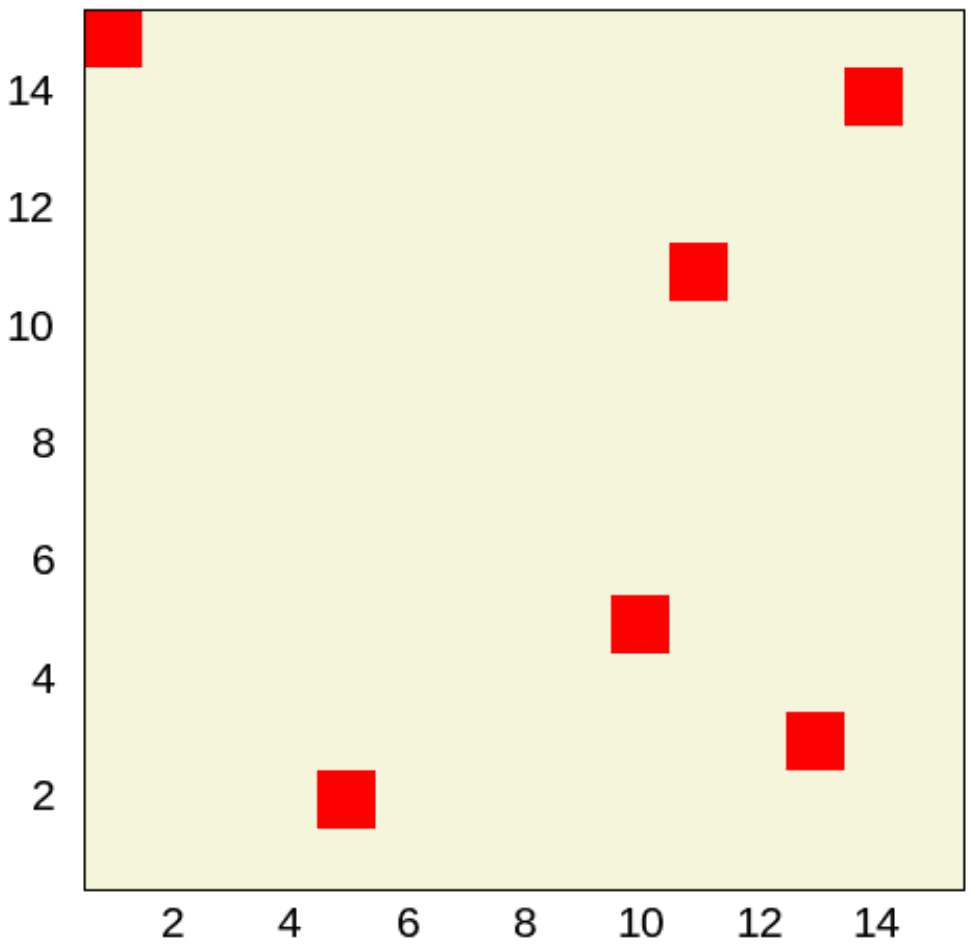}	
	\end{tabular}
	\caption{\small [From passive to active] Configurations of the model at different 
		times (increasing in lexicographic order) with
		parameters: $L=15$, $w_{\mathrm{ex}}=7$, and 
		$\varepsilon=0.3$. 
		Red pixels represent active particles, while blue pixels denote passive 
		particles, and gray sites are empty. 
		In the initial configuration (top left panel) there are $90$ active and 
		$90$ passive particles.}
	\label{fig:fig1}
\end{figure}

In the current version of the algorithm, we consider the communication between active and passive human groups by using the same techniques as in  \cite{Thieu2021}. In particular, each site on the lattice is connected by edges and the connectivity between these edges at one site is known as the degree of the site denoted by $\textrm{deg}(i,j)$, see Fig. \ref{fig:fig0-1}. The last part of the algorithm operates as follows. If an arrival site $(i,j)$ on the lattice is occupied by a passive particle, then we count the total number of active particles at its nearest neighbor sites. If the total number of actives particles is equal to $\textrm{deg}(i,j)-1$ at its nearest neighbor sites, then the passive particle is switched to the active one. Note that an arrival site of a site $x$ here is the random site chosen among possible empty sites at nearest neighbor sites of $x$. We expect that after communicating between active-passive human groups, almost all passive particles become active. Then, this seems to help the whole population in the room escape faster. The overall picture of our human crowd dynamics model is illustrated in Fig. \ref{fig:fig1}, where we show the configurations of the system at different times. 

\subsection{The evacuation time}

Considering the dynamics defined in Section~\ref{model},
we define $\tau_\eta$ as the following first 
hitting time to the empty configuration (e.g. \cite{Cirillo2019}):
\begin{equation}
	\label{eva000}
	\tau_\eta
	=
	\inf\{t>0:\,\eta(t)=\underline{0}\},
\end{equation}
for the chain started at $\eta \in \Omega$. 
Moreover, given a configuration $\eta\in\Omega$, the
\emph{evacuation time} starting from $\eta$ is defined as the following formulation
\begin{equation}
	\label{eva020}
	T_\eta
	=
	\mathbb{E}_\eta[\tau_\eta]
	\;.
\end{equation}
Here, the probability measure is induced by the Markov chain, described in Section \ref{model}, and $\mathbb{E}_\eta$ is the corresponding expectation. Note that in the limit, one can introduce the infinitesimal generator
$\mathcal{L}(\eta)
=\sum_{\eta'\in\Omega}
c(\eta,\eta')[f(\eta')-f(\eta)]$
acting on continuous bounded functions $f: \Omega \to \mathbb{R}$, see, e.g. \cite{Pavliotis2014,Spohn1991}.

According to \eqref{eva020}, we define the evacuation time here as the time 
needed to evacuate all the particles initially in the system. In other words, the evacuation time is the time at which the last particle leaves the 
room.
Next, we investigate 
the evacuation time of our human groups for a fixed initial 
random condition with specific values of the initial drift $\varepsilon$ and visibility depth 
$L_{\text{v}}$.
\begin{figure}[h!]
	\centering
	\begin{tabular}{ll}
		\includegraphics[width=0.45\textwidth]{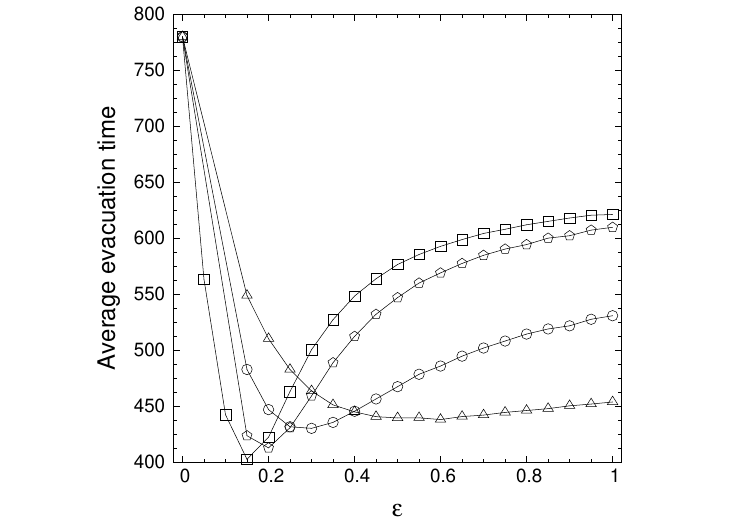}&
		\includegraphics[width=0.45\textwidth]{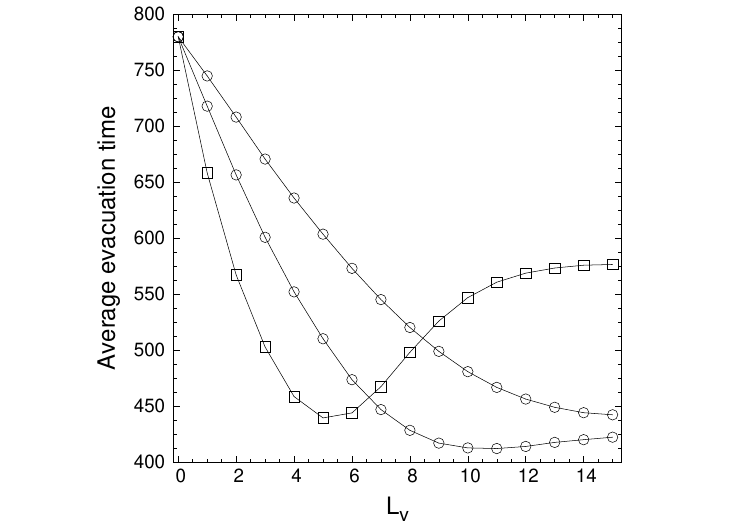}
	\end{tabular}
	\caption{Evacuation time of active-passive human groups. Left panel: evacuation time of active-passive particles as a function of $\varepsilon$ with $L_{\text{v}} = 5$ (open triangles), $L_{\text{v}} = 7$ (open circles), $L_{\text{v}} = 10$ (open pentagons), $L_{\text{v}} = 15$ (open squares). Right panel: evacuation time as a function of $L_{\text{v}}$ with $\varepsilon = 0.1$ (open triangles), $\varepsilon = 0.15$ (open circles), $\varepsilon = 0.2$ (open pentagons), $\varepsilon = 0.5$ (open squares).} 
	\label{fig:fig4}
\end{figure}

In the simulations, we fix the parameters for $L=15$, $w_{\mathrm{ex}}=\omega=7$ and $N_A=N_P=90$. The featured numerical results of our analysis are shown in Fig. \ref{fig:fig4}, where we have plotted the evacuation time as a function of $\varepsilon$ and $L_{\text{v}}$. These numerical results demonstrate that too \textquotedblleft smart\textquotedblright \ humans increase the evacuation time. In the left panel of Fig. \ref{fig:fig4}, for $L_{\text{v}}=5$, when we increase the values of $\varepsilon$, the evacuation time decreases up to $\varepsilon \approx 0.45$, then stays the same and slightly increases from $\varepsilon =0.8$ to $1$. For $L_{\text{v}}=7, 10, 15$, we note that the evacuation time reduces with $\varepsilon$ up to some values where it attains a minimum. However, soon after the evacuation time reaches its minimum, it increases significantly with the position and the size of the minimum change, respectively. This behavior is interesting since we expect that after the communication is included into consideration, the whole population should be able to escape faster. However, the presence of such a large visibility depth and drift quantity can increase the evacuation time of the whole population. This is due to the fact that after the communication between active and passive humans takes place, almost all passive humans become active humans. Moreover, all active humans are subject to a drift, guiding the active human group toward the exit door. Therefore, there is an accumulation of humans at the exit door that increases the evacuation time. This phenomenon is also demonstrated by the right panel of Fig. \ref{fig:fig4}. The numerical results that we observe here are a consequence of what is known as the \textquotedblleft faster-is-slower\textquotedblright \ effect. Normally, the \textquotedblleft faster-is-slower\textquotedblright \ effect is caused by the impatience of people in a panic mood. When people try to move faster, their average evacuation time increases. The accumulation of humans can cause clogging at an exit or a bottleneck that could lead to fatal accidents. This effect can be particularly tragic in
the presence of fires, where fleeing people can reduce their own chances of survival \cite{Helbing2005}.

As the next step, it is worth analyzing our system at the stationary state, which should provide further insight into the behavior we consider here.  
\subsection{The stationary occupation number profiles}

\begin{figure}[h!]
	\centering
	\includegraphics[width=0.55\textwidth]{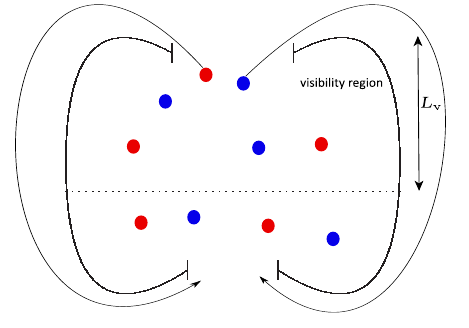}
	\caption{(Color online) Qualitative description of our model: red dots represent active particles, blue dots represent passive particles. Active particles are subject to a drift, guiding particles toward the exit door in the visibility zone. Outside the visibility region all particles are moving isotropically. All particles leaving the room via the top exit door are introduced back to the system via an entrance door at the bottom of the geometry.}
	\label{fig:fig5}
\end{figure}

Stationary crowd groups play an important role in studying the influences of pedestrian walking patterns \cite{Yi2015}. There are many applications of such stationary crowd groups, including walking path prediction, destination prediction, personality prediction, abnormal even detection, and self-organization phenomena \cite{Kok2016,Yi2014,Yi2015,Cirillo2017}. Moreover, from the statistical mechanics perspective, it is worth looking at the stationary state to detect non-trivial behaviors as time elapses beyond a characteristic walking timescale. When the system reaches a final stationary state, such a state can measure the outgoing fluxes of active-passive human groups.  
In this subsection, we examine the stationary state of our model following the ideas of an analysis discussed in \cite{Cirillo2020}. In particular, we modify the model presented in Section \ref{model} by considering the room with two doors, top and bottom. The bottom door, located at the bottom row of the lattice $\Lambda$, is symmetric with respect to its median column. It has the same width as the top door. We assume that every \textquotedblleft particle\textquotedblright \ exiting 
the domain via the top door is introduced back at one site randomly 
chosen among possible empty sites of the bottom door, see Fig. \ref{fig:fig5}. Then, the system will reach a stationary state.
Note that there is a slight fluctuation in the total number of active and passive particles
due to the fact that particles 
may enter waiting lists during evolution. Moreover, the total number 
of particles $N$ 
in the system is conserved by considering both the room and the 
waiting lists.

We shall consider
\emph{occupation number profiles} of active and passive particles, i.e. we evaluate the stationary 
mean value of the occupation numbers of active and passive particles.
 We assume that active-passive particles move inside the lattice $\Lambda$ with the following rates:
\begin{itemize}
	\item[$\bullet$]
	a passive particle leaves the room from a site in the top door 
	with rate $1$;
	\item[$\bullet$] 
	an active particle leaves the room from a site in the top door 
	with rate $1+\epsilon(x,y)$;
	\item[$\bullet$]
	if $n_{\textrm{A}} < N_{\textrm{A}}$ and 
	 $m_\textup{B} \neq 0$,
	then an active particle is added to a randomly chosen 
	empty site of the bottom door with rate 
	$[N_{\textrm{A}}-n_{\textrm{A}}]/m_\textup{B}$;
	\item[$\bullet$]
	if $n_{\textrm{P}} < N_{\textrm{P}}$ and $m_\textup{B} \neq 0$,
	then a passive particle is added to a randomly chosen 
	empty site of the bottom door with rate 
	$[N_{\textrm{P}}-n_{\textrm{P}}]/m_\textup{B}$;
	\item[$\bullet$]
	a passive particle moves inside $\Lambda$
	from a site to one of its empty nearest neighbors 
	with rate $1$; 
	\item[$\bullet$]
	an active particle 
	moves inside $\Lambda$ from a site to one of its empty nearest 
	site
	with rate $1+\epsilon(x,y)$.
\end{itemize}

Here, $n_{\textrm{A}}$ and $n_{\textrm{P}}$ denote the numbers of active and passive particles in the room, respectively. The numbers of active and passive 
particles that exited the room and entered their own waiting lists at the 
considered time are represented by
the quantities 
$N_{\textrm{A}}-n_{\textrm{A}}$
and 
$N_{\textrm{P}}-n_{\textrm{P}}$, respectively. Moreover, $m_\textup{B}>0$
is the number of empty sites of the bottom door at the same time.

In the study of these dynamic processes, one of the main quantities of interest is the 
stationary \emph{outgoing flux} or \emph{current	of active and passive particles}. The current is defined in the infinite time limit by the
ratio between the total number of active and passive particles, 
that in the interval $(0,t)$
exited through the top door 
to enter the waiting lists, and the time $t$.
In order to better understand the behavior of currents with 
respect to the model parameters, we shall also look at the 
\emph{occupation number profiles} of active and passive 
particles. Using the same methodology as in \cite{Cirillo2020}, we will evaluate the stationary 
mean value of the occupation numbers of active and passive 
particles.

\begin{figure}[h!]
	\centering
	\begin{tabular}{ll}
		\includegraphics[width=0.45\textwidth]{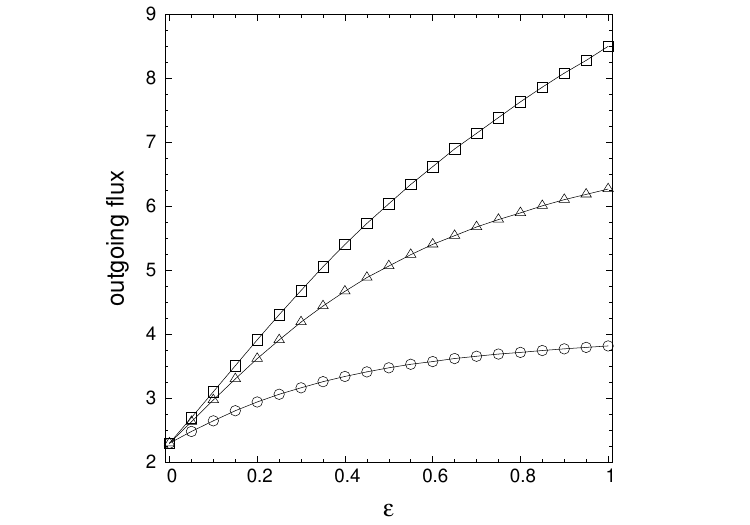}&
		\includegraphics[width=0.45\textwidth]{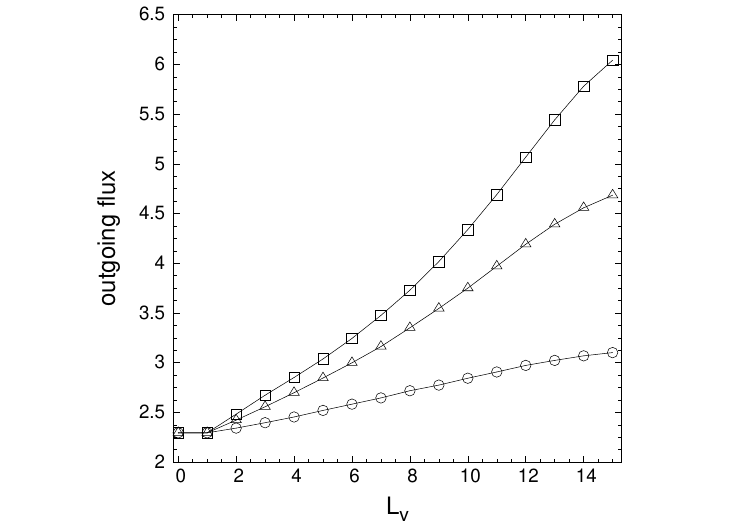}
	\end{tabular}
	\caption{Stationary current for the case of communication between active-passive populations. Left panel: $L_{\text{v}}=7$ (empty circles), $L_{\text{v}}=12$ (empty triangles), $L_{\text{v}}=15$ (empty squares). Right panel: $\varepsilon=0.1$ (empty circles), $\varepsilon=0.3$ (empty triangles), $\varepsilon=0.5$ (empty squares)}
	\label{fig:fig6}
\end{figure}

\begin{figure}[h!]
	\centering
	\begin{tabular}{lll}
		\includegraphics[width=0.3\textwidth]{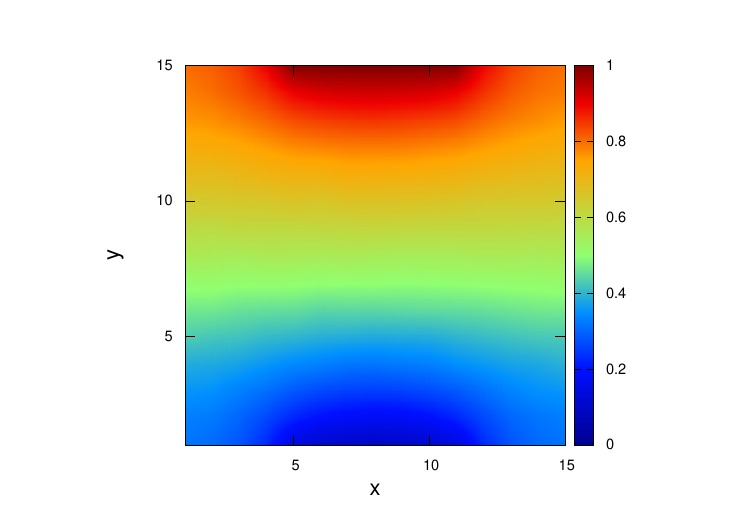}&
		\includegraphics[width=0.3\textwidth]{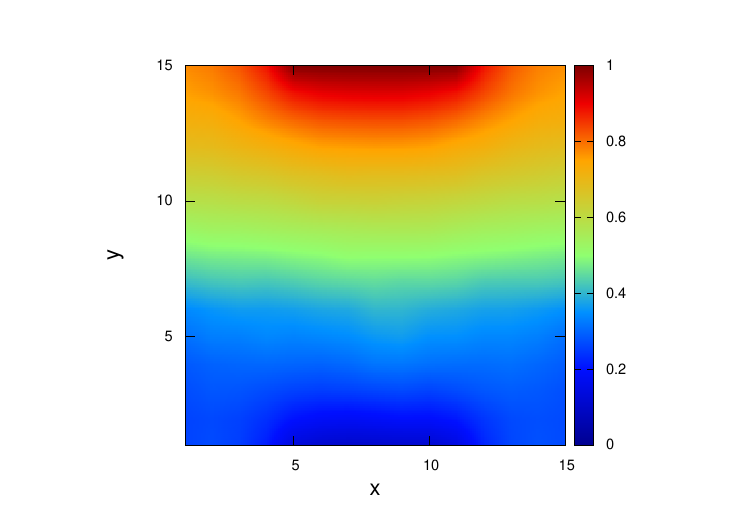}&
		\includegraphics[width=0.3\textwidth]{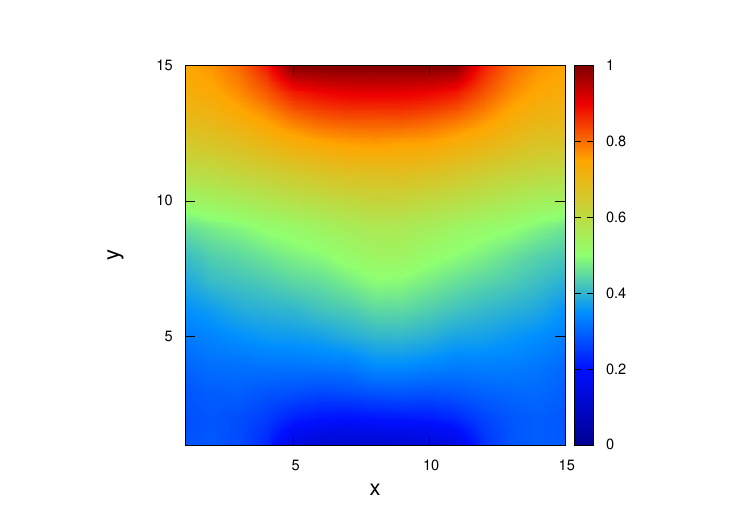}\\[0.1cm]
		\includegraphics[width=0.3\textwidth]{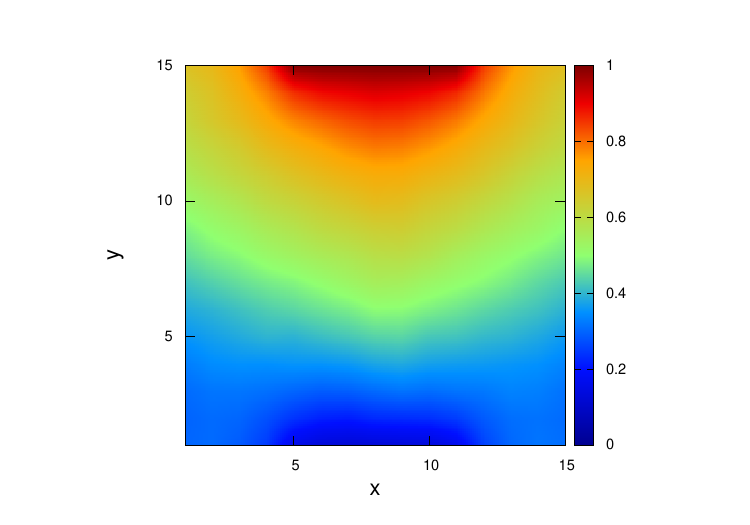}&
		\includegraphics[width=0.3\textwidth]{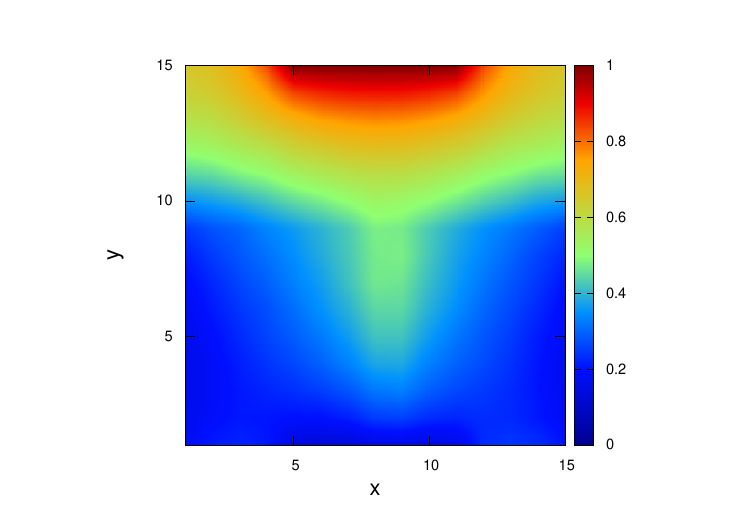}&
		\includegraphics[width=0.3\textwidth]{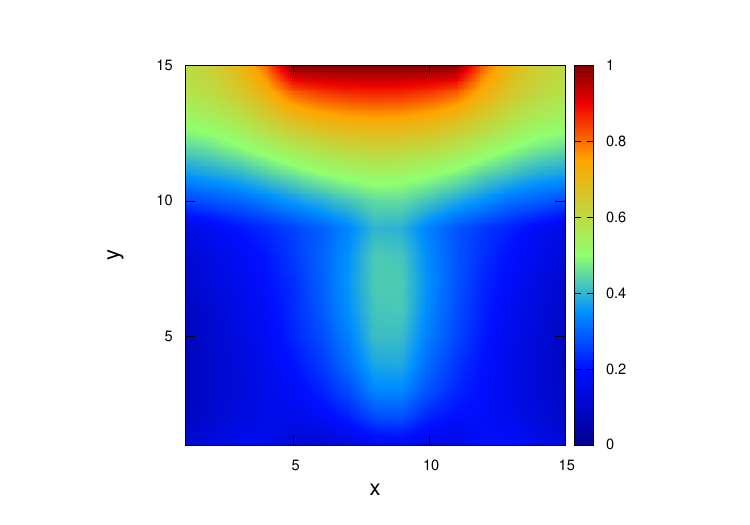}
	\end{tabular}
	\caption{\small Heat maps (different values of $\varepsilon$) for the case of interaction between active-passive humans: $L_{\text{v}}=7$ and $\varepsilon=0$, $L_{\text{v}}=7$ and $\varepsilon=0.1$, $L_{\text{v}}=10$ and $\varepsilon=0.1$, $L_{\text{v}}=15$ and $\varepsilon=0.1$, $L_{\text{v}}=10$ and $\varepsilon=0.3$, $L_{\text{v}}=10$ and $\varepsilon=0.5$.}
	\label{fig:fig7}
\end{figure}

\begin{figure}[h!]
	\centering
	\begin{tabular}{lll}
		\includegraphics[width=0.3\textwidth]{hmn4-00-7-AU-UA.pdf}&
		\includegraphics[width=0.3\textwidth]{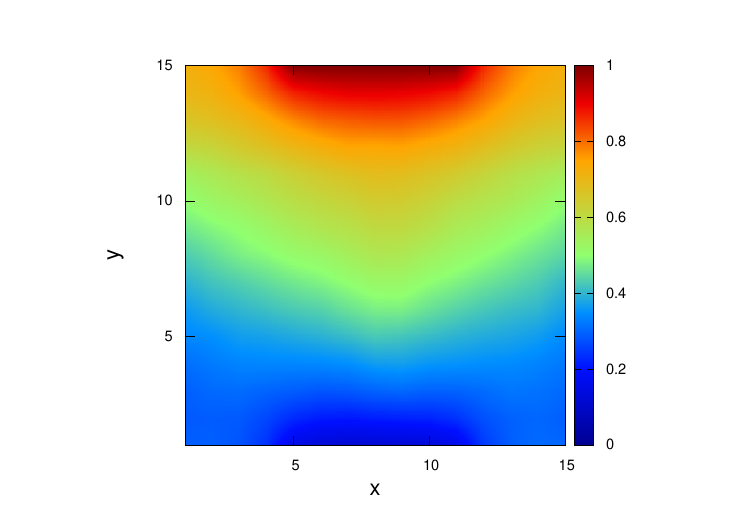}&
		\includegraphics[width=0.3\textwidth]{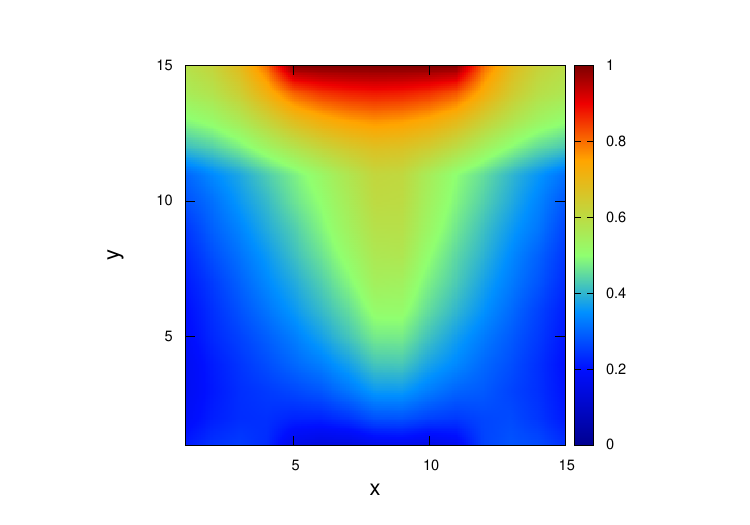}\\[0.1cm]
		\includegraphics[width=0.3\textwidth]{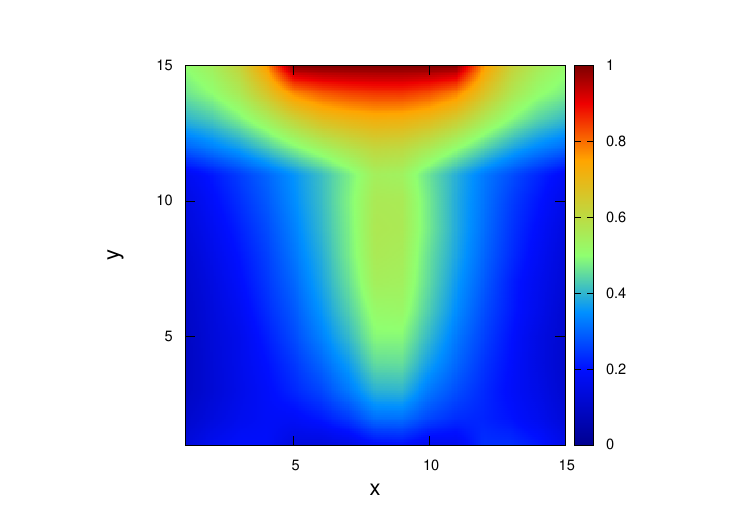}&
		\includegraphics[width=0.3\textwidth]{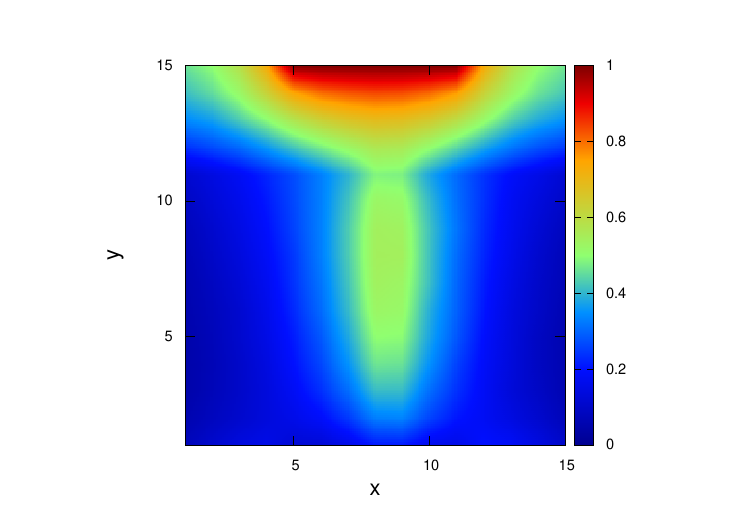}&
		\includegraphics[width=0.3\textwidth]{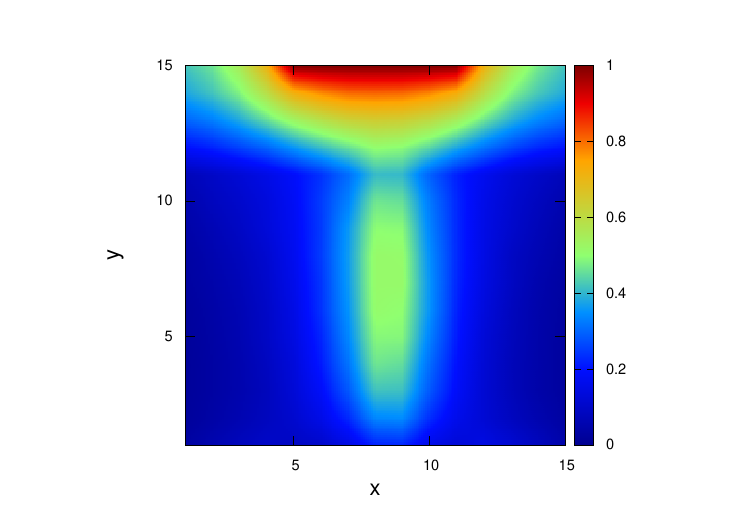}
	\end{tabular}
	\caption{\small Heat maps (different values of $\varepsilon$) for the case of interaction between active-passive humans $L_{\text{v}}=12$.}
	\label{fig:fig8}
\end{figure}

\begin{figure}[h!]
	\centering
	\begin{tabular}{lll}
		\includegraphics[width=0.3\textwidth]{hmn4-00-7-AU-UA.pdf}&
		\includegraphics[width=0.3\textwidth]{hmn4-01-15-AU-UA.pdf}&
		\includegraphics[width=0.3\textwidth]{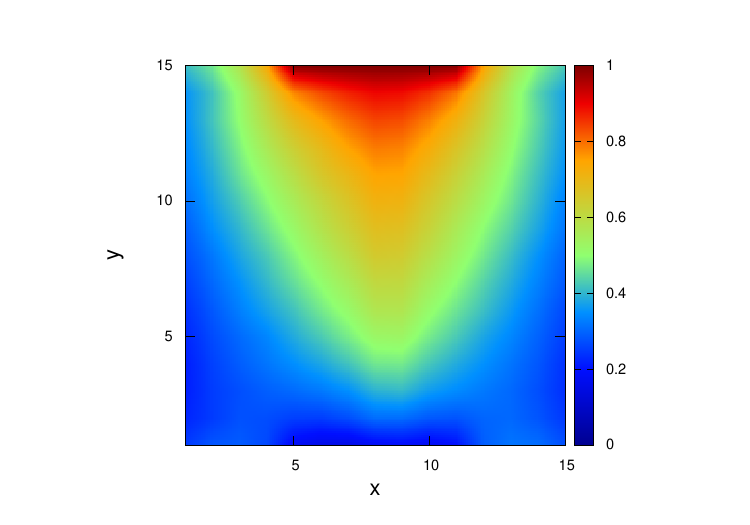}\\[0.1cm]
		\includegraphics[width=0.3\textwidth]{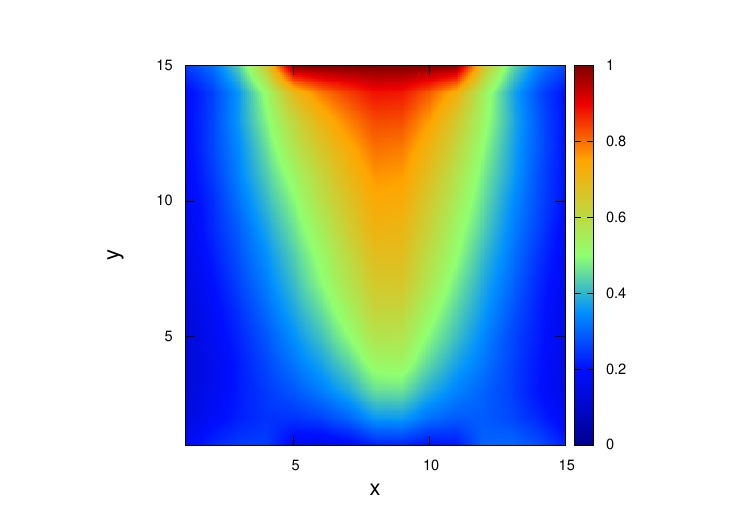}&
		\includegraphics[width=0.3\textwidth]{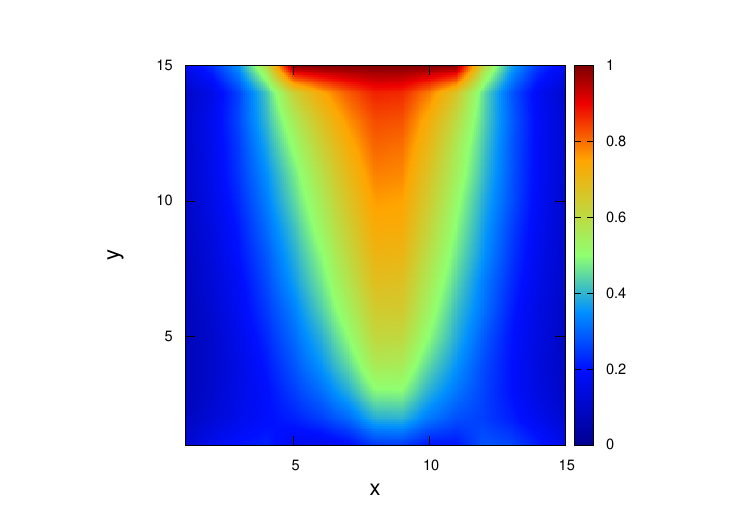}&
		\includegraphics[width=0.3\textwidth]{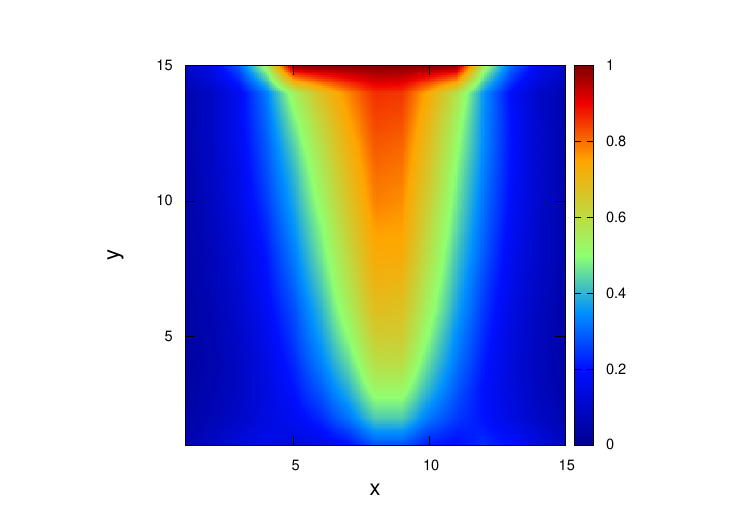}
	\end{tabular}
	\caption{\small Heat maps (different values of $\varepsilon$) for the case of interaction between active-passive humans $L_{\text{v}}=15$.}
	\label{fig:fig9}
\end{figure}

In the simulations, we fix the parameters for $L=15$, $w_{\mathrm{ex}}=\omega=7$ and $N_A=N_P=90$. The featured numerical results of our analysis here are shown in Figs. \ref{fig:fig6}-\ref{fig:fig9}. In Fig. \ref{fig:fig6}, we plotted the stationary currents as functions of $\varepsilon$ and $L_{\text{v}}$. We observe that the behavior of the currents of active-passive humans demonstrates an expected pattern, in which the currents increase monotonically both with respect to the drift quantity and the length of the visibility region. In particular, in the left panel of Fig. \ref{fig:fig6} , when we increase the values of $\varepsilon$, the currents of active-passive humans increase significantly. Moreover, the current of active-passive humans in the case of $L_{\text{v}}=7$ is smaller than in the cases of $L_{\text{v}}=12$ and $L_{\text{v}}=15$. This is also evident from the right panel, where the currents increase when we increase the values of  $L_{\text{v}}$.

In order to analyze this effect further, we look at the corresponding occupation number profile, obtained from our model. In what follows, we use a data visualization technique known as heat maps \cite{Wilkinson2009}. As seen from Figs. \ref{fig:fig7} - \ref{fig:fig9}, active humans are mainly distributed at their exit door. This is observable in all of the panels in these figures. For large values of $\varepsilon$ and $L_{\text{v}}$, due to the transverse component 
of the drift, there is an accumulation of particles in 
the central part of the room. Moreover, it is clear that particles can form a central \textquotedblleft droplet\textquotedblright \  detached 
from the \textquotedblleft inlet\textquotedblright \  top door depending on the length of the visibility region.  

\subsection{Comparison with the case without interaction between active-passive particles}

In this section, we compare the evacuation times in two different situations: with and without communication between active and passive humans. 
\begin{figure}[h!]
	\centering
	\begin{tabular}{ll}
		\includegraphics[width=0.45\textwidth]{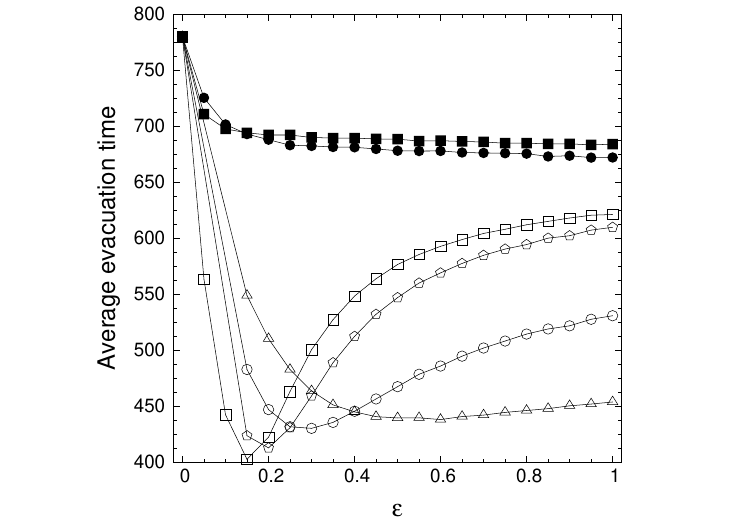}&
		\includegraphics[width=0.45\textwidth]{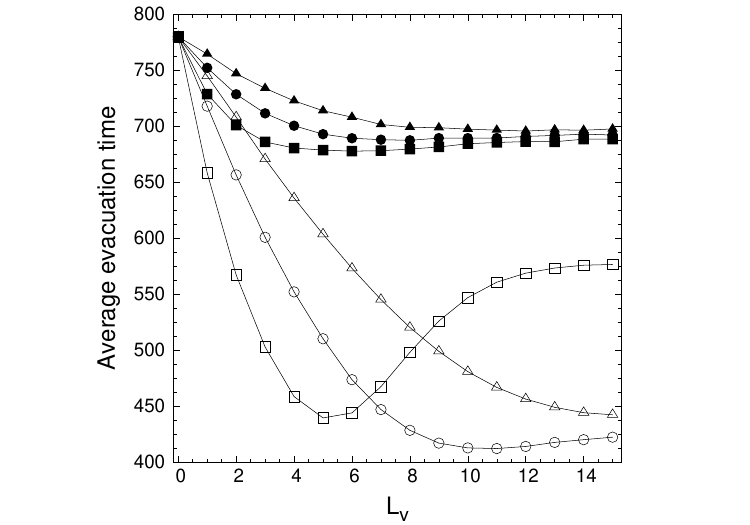}
	\end{tabular}
	\caption{Evacuation time of active-passive human groups: no communication between active-passive human groups (solid disks), with communication between active-passive human groups (open symbols). Left panel: evacuation time of active-passive particles as a function of $\varepsilon$ with $L_{\text{v}} = 5$ (triangles), $L_{\text{v}} = 7$ (circles), $L_{\text{v}} = 10$ (pentagons), $L_{\text{v}} = 15$ (squares). Right panel: evacuation time as a function of $L_{\text{v}}$ with $\varepsilon = 0.1$ (triangles), $\varepsilon = 0.2$ (circles), $\varepsilon = 0.5$ (squares).} 
	\label{fig:fig4-2}
\end{figure}
In the simulations, we fix the parameters for $L=15$, $w_{\mathrm{ex}}=\omega=7$ and $N_A=N_P=90$. The featured numerical results from our analysis in this subsection are shown in Fig. \ref{fig:fig4-2}, where we have plotted the evacuation time as a function of $\varepsilon$ and $L_{\text{v}}$ for the two considered cases, with and without communication between active-passive humans. In particular, in the left panel of Fig. \ref{fig:fig4-2}, for $L_{\text{v}}=7$ and $L_{\text{v}}=15$, the average evacuation time without communication between active and passive humans is larger than the average evacuation time with communication between the two human groups when we increase the value of $\varepsilon$. Moreover, it is clear that even the average evacuation time in the dynamics with communication between active-passive humans reduces significantly up to some minimum values and then increases dramatically under suitable values of $\varepsilon$ and $L_{\text{v}}$. The average evacuation time with communication between active-passive humans is still smaller than in the case without communication between the two groups. Hence, the communication between active and passive humans strongly affects the evacuation time of the whole populations. This is visible also in the right panel, for $\varepsilon=0.1, 0.2, 0.5$, when we increase the values of $L_{\text{v}}$. Indeed, the evacuation time in the dynamics of exchanging information between active and passive humans is smaller than in the case without communication between the two human groups. Even for large enough value of $\varepsilon=0.5$ and $L_{\text{v}}$ from $5$ to $15$, the evacuation time in the dynamics of switching from passive to active humans is still smaller than in the case when there is no exchange of the information between active and passive populations. Finally, it is worth noting that the communication between active and passive humans strongly affects the evacuation time of the whole population even in the presence of a classical \textquotedblleft faster-is-slower\textquotedblright \ phenomenon. Note also that in the dynamics without communication, there is the presence of only a simple exclusion constraint of the lattice gas dynamics. This special case has been reported in \cite{Cirillo2019}, and our numerical results can be compared with the numerical results presented there. Using a similar lattice gas model with two species of active and passive particles, the authors in \cite{Cirillo2019} have studied the pedestrian escape from an obscure room. Their main observable is the evacuation time of pedestrians. They have found that the presence of active pedestrians favors the evacuation of the passive ones even if no information exchange is allowed, and even in the presence of the exclusion constraint of the lattice gas dynamics. However, as an extension of the model provided in \cite{Cirillo2019}, our model investigates the exchange information of the geometry from active pedestrians to passive pedestrians. We found that the communication between active and passive pedestrians dramatically improves the evacuation time of the whole population, even in the presence of a classical \textquotedblleft faster-is-slower\textquotedblright, see, e.g. Fig. \ref{fig:fig4-2}.

\begin{figure}[h!]
	\centering
	\begin{tabular}{ll}
		\includegraphics[width=0.45\textwidth]{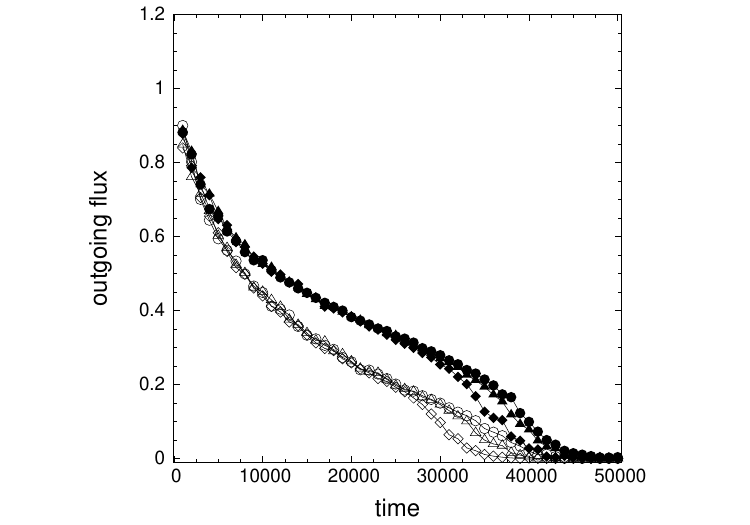}&
		\includegraphics[width=0.45\textwidth]{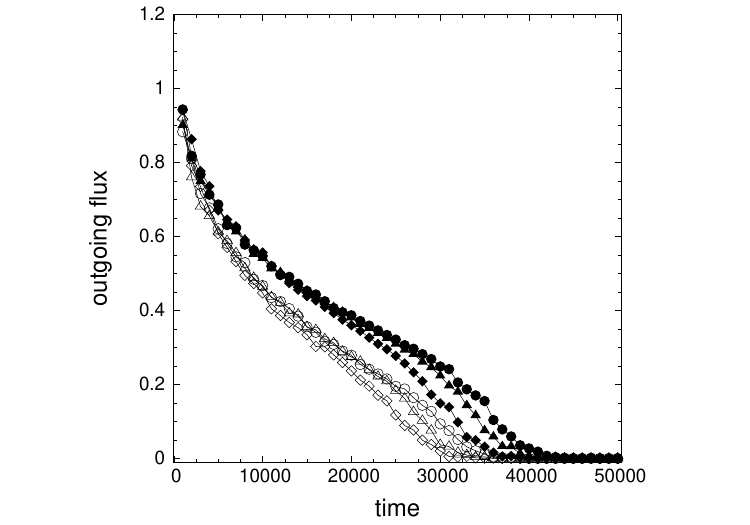}\\[0.1cm]
		\includegraphics[width=0.45\textwidth]{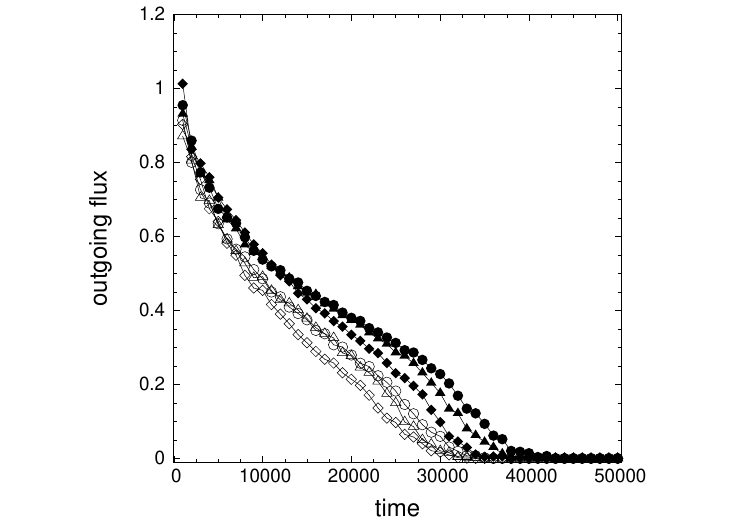}&
		\includegraphics[width=0.45\textwidth]{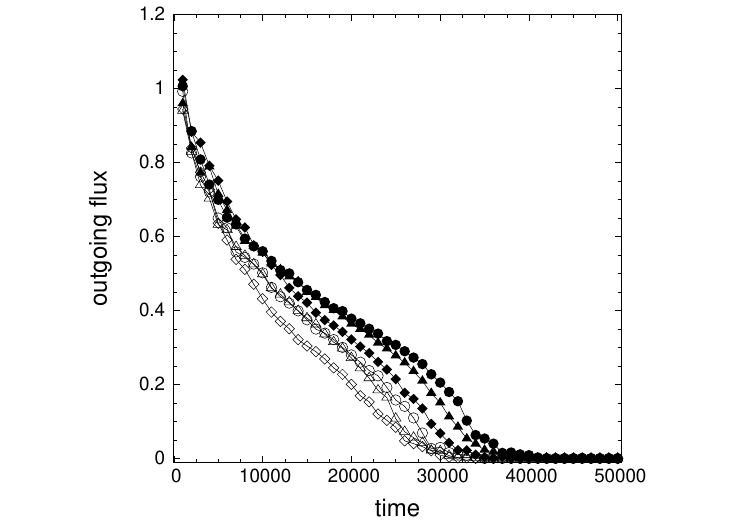}
	\end{tabular}
	\caption{Transient flux of active-passive human groups: no communication between active-passive human groups (solid disks), with communication between active-passive human groups (open symbols). Parameters: with $L_{\text{v}} = 5$ (circles), $L_{\text{v}} = 7$ (triangles), $L_{\text{v}} = 15$ (squares). Top left panel: $\varepsilon = 0.1$. Top right panel:  $\varepsilon = 0.3$. Bottom left panel: $\varepsilon = 0.5$. Bottom right panel: $\varepsilon = 0.7$.} 
	\label{fig:fig4-3}
\end{figure}
We also look at the behavior of the transient fluxes as functions of time with the same setting discussed in Section 2 (see, e.g., \cite{Cirillo2019}). However, in this transient fluxes calculation, no extra door at the bottom is included, i.e. every human who exists in the domain via the top door is not introduced back at the bottom door. In order to do that, we averaged the flux of passive humans over 103 different realizations of the process for a given initial condition. In Fig. \ref{fig:fig4-3}, we compare the results of transient fluxes of active-passive human groups between no communication between active-passive human groups and communication between active-passive human groups. We observe that the communication between active-passive human groups enhances the outgoing flux of all populations. This result is again a consequence of the underlying benefits of transferring information between active-passive human groups during evacuation.
		 The communication between active-passive human groups also aligns with the so-called "leader-and-follower" phenomena \cite{Zhang2021}, where the active human group can be considered as leaders, and the passive human group can be represented as followers.

\section{Conclusions and future directions}
We have proposed and described a statistical-mechanics-based lattice gas model for active-passive populations, focusing on the application of human crowd behaviors in critical situations. Specifically, we have deduced the average evacuation time of active-passive human groups, where we have observed a classical \textquotedblleft faster-is-slower\textquotedblright \ effect. Our numerical results have shown that the communication between active and passive human groups strongly influences the evacuation time of the whole population even in the presence of a standard \textquotedblleft faster-is-slower\textquotedblright \ phenomenon. Moreover, we have investigated the  stationary state of our model. In particular, the numerical results have demonstrated that the drift quantity and the length of the visibility region strongly affect the shape of the region where active humans accumulate. In general, the \textquotedblleft faster-is-slower\textquotedblright \ effect and clogging at the narrow exits are not well understood due to their real-world complexity, and this contribution sheds light on these important issues. As a continuation of this work, it would be instructive to further investigate such bottleneck situations during emergency, since the instant herding behaviour of humans at the exit could lead to fatal accidents. We note that the ideas presented in this contribution may be extended to a system of coupled Langevin equations to observe the detailed dynamics of active-passive particles at the exit door.

Given that the main exemplifications here were given for critical situations such as possible
disasters and cases of emergency requiring evacuations, an important extension of the present
work would also be to account for psychological effects such as emotions (e.g., induced by
fear). Some works in this direction have already been carried out \cite{Xu2021,Saha2023,Ren2023,Van2023}.
Our final remark goes to developing intelligent transportation systems in smart cities \cite{Melnik2009,Wang2023} that require coupled modelling of human crowds and traffic flows.
In such situations, models for pedestrian-vehicle mixed traffic flows become a building block
for smart city designs where cellular automata models can provide an appropriate framework,
as recent papers demonstrate (\cite{Wang2023} and references therein). In this area, data-driven
modelling accounting for human factors, crowd behaviour optimization, and ultimately
control of such coupled systems involving human crowds become a cornerstone of the
success of the entire design process. At the same time, challenges in the application of LGCA
methods are well-known, which, apart from the lack of Galilean invariance and difficulties of
extensions to three-dimensional problems, include statistical noise and incompleteness of the
information about the system. In applying to multiscale problems of crowd dynamics, the
latter challenge may be addressed with the availability of additional data from the
microscopic scale. In Section 2, we mentioned that this would require learning the
Hamiltonian of many-body systems of interacting particles, a notoriously difficult task if
Molecular Dynamics simulations are to be used. Nevertheless, recently it was shown that this
task can potentially be carried out even for more complex Hamiltonians that go beyond the
realm of our problems here once such tools as those based on learning Boltzmann machines
(BMs) or Restricted Boltzmann machines (RBMs) are used \cite{Anshu2021,Gebhart2023}. Boltzmann machines
are usually viewed as unsupervised learning neural networks \cite{Tao2021}. In fact, a central problem
in machine learning is learning Boltzmann machines, which have played a significant role in
many application areas relevant to big data, including deep learning architectures. From a
physics point of view, a BM is an Ising model, which is a stochastic spin-glass model with an
external field. Since Hamiltonians of spin glasses are used as a starting point to define the
learning task, such models are often termed in machine learning and cognitive sciences as
energy-based models. A RBM is a restricted stochastic Ising model based on a generative
stochastic artificial neural network with the restriction on their neurons (they must form a
bipartite graph) that can learn a probability distribution over its set of inputs. While BMs
allow connections between hidden layer units, RBMs have no connections between nodes
within a group, which can lead to more efficient training algorithms. As we mentioned above,
data-driven modelling with subsequent crowd behaviour optimization and control is vital for
designing advanced intelligent transportation systems for smart cities, and cellular automata
models lead the way in many of these new developments. At the bigger picture, such models
are discrete models of computation that are studied in automata theory and mathematical logic,
with applications to many fields of science and engineering, including the theory of
computation and Artificial Intelligence. Along with the growing application of BMs and other
neural-network-based models in the data-driven modelling of human crowds \cite{Zhai2021,He2023,Luca2021,Bahamid2022}, the
LGCA models developed here and their further extensions provide a simple, viable, and
efficient tool for new advances in these areas.
\section*{Acknowledgments}

	Authors are grateful to the NSERC and the CRC Program for their
	support. RM is also acknowledging support of the BERC 2022-2025 program and Spanish Ministry of Science, Innovation and Universities through the Agencia Estatal de Investigacion (AEI) BCAM Severo Ochoa excellence accreditation SEV-2017-0718 and the Basque Government fund AI in BCAM EXP. 2019/00432.

\bibliography{mybibn-old}
%

\end{document}